\documentclass[prb,byrevtex,floatfix,twocolumn,superscriptaddress]{revtex4}
\usepackage{amssymb,amsmath}
\usepackage{graphicx,afterpage}
\newcommand{\minitab}[2][l]{\begin{tabular}{#1}#2\end{tabular}}
\usepackage{txfonts}
\usepackage{multirow}

\begin{document}

\title{Symmetry conditions for non-reciprocal light propagation in magnetic crystals}

\author{D\'avid Szaller}
\affiliation{Department of Physics, Budapest University of
Technology and Economics and Condensed Matter Research Group of the
Hungarian Academy of Sciences, 1111 Budapest, Hungary}

\author{S\'andor Bord\'acs}
\affiliation{Department of Physics, Budapest University of
Technology and Economics and Condensed Matter Research Group of the
Hungarian Academy of Sciences, 1111 Budapest, Hungary}
\affiliation{Quantum-Phase Electronics Center,
Department of Applied Physics, The University of Tokyo, Tokyo
113-8656, Japan}

\author{Istv\'an K\'ezsm\'arki}
\affiliation{Department of Physics, Budapest University of
Technology and Economics and Condensed Matter Research Group of the
Hungarian Academy of Sciences, 1111 Budapest, Hungary}

\date{\today}

\begin{abstract}
Recent studies demonstrated the violation of reciprocity in
optical processes in low-symmetry magnetic
crystals. In these crystals the speed of light can be different for counter-propagating beams. Correspondingly, they can show strong directional anisotropies such as direction dependent absorption also called directional dichroism[S. Bord\'acs et al., Nat. Phys. {\bf 8}, 734 (2012); M. Saito et al., J. Phys. Soc. Jpn. {\bf 77}, 013705 (2008)]. Based on symmetry considerations, we identify the magnetic point
groups of materials which can host such directional anisotropies and also provide a list of possible candidate materials to observe these phenomena. In most of these cases, the symmetry of the crystal
allows directional anisotropy not only for optical processes but also for
the propagation of beams of particles and scalar waves.
We also predict new types of directional optical anisotropies --
besides the optical magnetoelectric effect and the magnetochiral
dichroism investigated so far -- and specify the magnetic point
groups of crystals where they can emerge.
\end{abstract}

\maketitle

When a measured quantity of matter remains invariant under the
interchange of the source and the detector, the corresponding
symmetry of the studied system is termed as reciprocity. Reciprocity in physics has a challenging definition. While Landau used the term reciprocity as a synonym of time reversal symmetry \cite{Landau}, later it was generalized by also considering spatial symmetries of the studied system\cite{ShelankovPRB1992}. A rigorous definition of reciprocity for scattering processes has been recently given by De\'ak and F\"ul\"op\cite{Deak}.

Specific to optical phenomena, among reciprocity-violating media we consider those which can distinguish between counter-propagating light beams irrespective of the polarization state of the light, i.e. we say a material shows directional anisotropy when such non-reciprocity is present for each polarization of the light beam.
As a counter-example, we recall the well-known Faraday effect when the non-reciprocity holds only for circularly polarized
light\cite{Potton,Barron}, while counter-propagating linearly polarized beams have the same transmission and reflection properties. In this case the reciprocity is violated for certain but not all polarizations, thus, in our term the corresponding media show no directional anisotropy.

\begin{figure}[h!]
\includegraphics[trim=0cm 0cm 0cm 0cm, clip=true, width=3.4in]{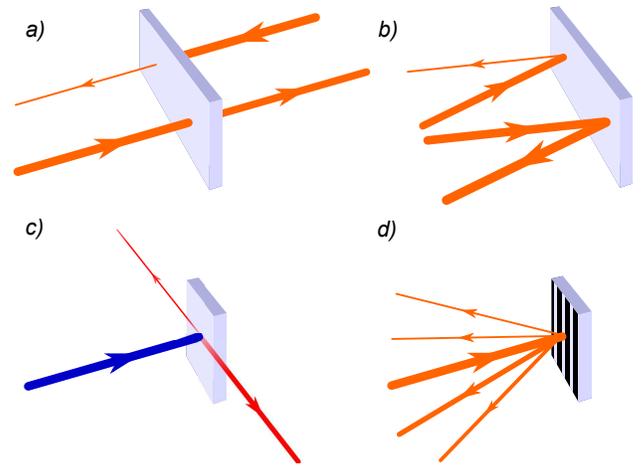}%
\caption{(Color online) Schematic representation of direction-dependent light propagation in {\it (a)} transmission, {\it (b)} reflection, {\it (c)} emission and {\it (d)} Bragg-scattering. In case of {\it (a)} transmission and {\it (b)} reflection, forward and backward propagating beams distinguished by the medium are simultaneously shown. {\it (c)} Excitations by higher energy photons (blue arrow) decay via emission characterized by direction-dependent emission rates (red arrows). {\it (d)} Equal intensity for beams diffracted symmetrically to the direct beam is broken by directional anisotropy.}
\label{fig1}
\end{figure}

Although the presence of such reciprocity-violating directional anisotropy in the properties of bulk matter may seem strange and counter-intuitive, it has been recently verified in a large variety of low-symmetry materials. In an absorbing medium this directional anisotropy results in different strengths of absorption for beams traveling along opposite directions which is called directional dichroism. Similar directional optical phenomena -- schematically shown in Fig.\ref{fig1} -- have already been observed for emission\cite{RikkenNat1997,RikkenB}, absorption\cite{Hopfield}
 and diffraction\cite{Koerdt} of light in materials belonging to two different symmetry classes where the directional optical phenomena are manifested in the corresponding two configurations.

The optical magnetoelectric effect (OME) was observed in systems with
finite ferrotoroidal moment\cite{Arima} -- which can arise in simple cases as the cross product of the static
magnetization and electric polarization of the material, $\boldsymbol{T=P\times M}$ -- for light
propagating along or opposite to the toroidal moment vector. Strong
OME was first detected by Hopfield and Thomas in the absorption
spectrum of excitons in the polar CdS with wurtzite structure
\cite{Hopfield}. Instead of reversing the light propagation, the
sign of the toroidal moment was switched to be parallel and
antiparallel to the velocity of light. A remarkable portion of the
absorption was found to be odd function of both the magnetization and
the electric polarization of the material as expected for OME. Later the same phenomenon
was found for the polar ferrimagnet GaFeO$_{3}$ in the X-ray absorption
of the core electron excitations \cite{Kubota2004} and for
absorption due to intra-atomic d-d transitions over the visible-ultraviolet spectral range\cite{Jung2004,Kida2006}. Enhancement of OME in diffraction was
observed using a grating made of the same compound\cite{Kida2006}. Metamaterials composed of two dimensional ferromagnetic islands with polar shape can also host OME in diffraction\cite{KidaPRL2005}.

Materials without spontaneous electric polarization and/or
magnetization can show OME as well if a toroidal moment $\boldsymbol{T}$ is induced by
crossed static electric and magnetic fields, as was observed for the
electronic f-f transitions of Er$^{3+}$ in
Er$_{1.5}$Y$_{1.5}$Al$_{5}$O$_{12}$ \cite{RikkenPRL2002}. Besides the usual case discussed above, spontaneous toroidal moment can also be present in antiferroelectric and/or antiferromagnetic compounds, whenever $\boldsymbol{t}=\sum_i \boldsymbol{r}_i\times \boldsymbol{m}_i$ is finite \cite{Spaldin2008}, while $\boldsymbol{T}$ can be vanishing. Here, $\boldsymbol{m}_i$ and $\boldsymbol{r}_i$ respectively stand for the $i$-th magnetic moment and its position. For example the spin texture present in skyrmion lattices\cite{Cu2OSeO3, FeGe, MnSi} can produce such toroidal moment.

Besides its
manifestation in optical processes, similar directional anisotropy has been reported in diffusive electronic transport. The
drain-source resistance of symmetric field effect transistors, when
applying crossed magnetic field and electric field (via a finite
gate-channel potential), showed difference for current flowing along
or opposite to the toroidal moment \cite{RikkenPRL2005}. Directional anisotropy was
predicted recently for the propagation of spin-waves in planar
magnonic crystals with metallic over- and dielectric underlayers
\cite{Krawczyk2012} and in an ultrathin Fe film on a W(110) surface
\cite{Szunyogh2009}. The latter was experimentally observed by
spin-polarized electron energy loss spectroscopy \cite{Zakeri2010}.
In both cases ferrotoroidal moment was introduced parallel to the wave propagation by crossed magnetic field or spontaneous magnetization and electric polarization arising from the layered structure.

Except for
the case of CdS, the previous examples for OME, detected in absorption/reflection/diffraction or in
electric resistance, gave small relative directional differences in
the range of $10^{-5}-10^{-2}$. Recent experiments on magnetoelectric multiferroic
materials show a relative OME in the order of unity as it was
demonstrated for intra-atomic d-d transitions of CuB$_2$O$_4$ in the near-infrared spectral region
\cite{SaitoJPSJ2008} and in the electromagnon spectrum of
Ba$_2$CoGe$_2$O$_7$ in the THz photon frequency range
\cite{Kezsmarki2011}, which may bring the possibility of their
optical applications e.g. as magnetically controlled directional
light switches.

The second type of optical directional anisotropy effect, the
magneto-chiral dichroism (MChD) appears as a difference in the
absorption or emission of chiral systems for light beams propagating
parallel and antiparallel to an external magnetic field. The
existence of MChD was first experimentally proved for a luminescent
transition of the chiral Eu(($\pm$)tfc)$_3$ complex
\cite{RikkenNat1997,RikkenB} where the intensity of the emitted
light slightly differed for propagation along or opposite to the
external magnetic field. The effect was also shown in the infrared
absorption of $\alpha - $NiSO$_4 \cdot 6 $H$_2$O \cite{RikkenPRE1998}
and in the enantioselective photochemical dissociation of chiral
complexes \cite{RikkenNat2000}. Similar phenomenon, difference in
phase retardation of counter-propagating light beams was found in
chiral organic liquids \cite{Kleindienst1998,Vallet2001}. The
emergence of MChD was also demonstrated in Bragg scattering
\cite{Koerdt}. The electrical analog of MChD is the difference in
conductivity of chiral systems for currents flowing parallel and
antiparallel to the direction of an external magnetic field. Such
electric magnetochiral anisotropy was observed in helically
distorted bismuth stripes \cite{RikkenPRL2001} and also in chiral carbon
nanotubes \cite{nanotube}. In all these cases the directional
anisotropy was rather weak corresponding to a relative difference of
$10^{-8} - 10^{-2}$ in the studied quantities. Similar to the case
of OME, multiferroic materials can show also a significant MChD
effect in both their infrared d-d excitations
\cite{SaitoPRL2008} and spin-wave absorption spectra
\cite{Bordacs2012}.

Motivated by the recently observed plethora of directional anisotropy effects and their potential in future applications using multiferroic compounds, we study the general symmetry conditions allowing directional anisotropy in optical phenomena. Our focus is on crystalline materials where we systematically identify those crystallographic magnetic point groups which can
support directional anisotropy effects. Besides the symmetry conditions where OME and MChD can be realized,
we found additional magnetic symmetry groups which can host new
types of directional anisotropy effects. These new directional effects are specific to crystalline media in contrast to OME and MChD which can also emerge in the optical properties of gases, liquids and amorphous solids.

When neglecting the weak interaction light-matter interaction possesses time reversal\cite{timeEM} ($\tau$) and
full $O(3)$ spatial symmetry, in other words transformation of the whole
"light source-sample-detector system" by $\tau \otimes O(3)$ symmetry
operations must not change the interaction process. Therefore, only the
crystallographic and magnetic structure of matter can give
constraints to the symmetry of the measured optical quantities. Using Neumann's principle these constrains can provide relations between the tensor elements describing the linear optical response such as transmittance and reflectivity matrices. These relations can be experimentally probed by changing the polarization or propagation direction of light. We focus on such cases when the equality of these tensor components for forward and backward propagation is broken by the low symmetry of the material. Thus, every symmetry elements, which could convert
any of the optical processes into their reversed pairs, need to be eliminated from the magnetic point group of such crystal.

According to our definition, directional anisotropy is allowed for a
given propagation direction in a material, if no symmetry operation of the crystal can transform a light beam (of any
polarization) traveling along this direction into a
counter-propagating one without changing its polarization state. This means that the path of the electric polarization vector determines the same curve in the laboratory frame upon forward and backward propagation. As a necessary condition, the material must lack spatial inversion and time reversal symmetries because they convert counter-propagating beams with the same linear polarization into each other. To systematically determine the conditions of directional anisotropy in a crystal we have to consider each element of the corresponding magnetic point group. Let
us take a beam with arbitrary elliptical polarization
propagating along the $z$-axis. Since the electric displacement and
magnetic flux density vectors ($\boldsymbol{D}$ and $\boldsymbol{B}$, respectively) lie in the
$xy$-plane we only need to treat those orthogonal transformations
which leave this plane invariant. In crystals these are the $n$-fold
proper ($n$) and improper ($\overline{n}$) rotations around the $z$-axis
 for $n = 1,2,3,4$ and $6$ and around an axis
laying in the $xy$-plane for $n = 1$ and $2$. All of
these operations can be combined with the reversal of time which
will be denoted by a prime ($^{\prime}$). In this paper an improper
rotation means a proper rotation combined with spatial inversion, giving
$\overline{1}$ for the spatial inversion and $\overline{2}$ for a
mirror plane symmetry.

As long as we consider the propagation of unpolarized light, the
necessary and sufficient condition for the presence of a directional
anisotropy effect along the $z$-axis is that no symmetry of the
system can connect the opposite propagation directions, i.e.
transform light $\bf{k} \uparrow\uparrow z$ vector to $\bf{-k}
\downarrow\uparrow z$. Half of the transformations listed in the
previous paragraph fulfill this condition, since we have the freedom
to combine a spatial operation with time reversal, which converts a
propagation-reversing symmetry to a propagation-conserving one and
vice versa. Proper rotations ($n$) and improper rotations combined
with time reversal ($\overline{n}^\prime$) around the $z$-axis meet
this criterion for any $n \in \left\{ 1,2,3,4,6 \right\}$. Besides
that $\overline{2}$ and $2^\prime$ rotations around an axis
perpendicular to the $z$-direction also conserve the propagation
direction. In magnetic point groups containing only these
propagation-conserving operations -- together with any other
transformations which do not leave $xy$-plane invariant -- the
propagation along the $+z$ and $-z$ directions can be different not
only for unpolarized light but for other unpolarized or scalar waves
(e.g. acoustic waves) and for unpolarized beams of particles.

Next, we also take the polarization state of light into account, because there can be some propagation-reversing operations which also alter the polarization state of light (for each polarization) and thus allow directional optical anisotropy.
However, this will only slightly affect the results obtained above. Among the rotations
around the $z$-axis only $\overline{n}$ and $n^{\prime}$
reverse the propagation of light. Both
transformations rotate the axes of the polarization ellipse by $\frac{2 \pi}{n}$. While $n^{\prime}$ transposes clockwise and counterclockwise circulation of the polarization, $\overline{n}$ preserves the sense of circulation in the laboratory frame.
Since the direction of propagation is reversed, $n^{\prime}$ will preserve and
$\overline{n}$ will change the the sense of circulation in the frame of the
light beam. For $n = 1$ and $2$, $n^{\prime}$ connects
every elliptical polarization state with its counter-propagating
pair, and $\overline{n}$ does the same for linearly polarized states. Thus, both prohibit directional optical anisotropy (in our term) as at least for specific polarizations the propagation direction is reversed and the polarization in the light frame is kept intact. For $n = 5$ the $n$-times repetition of $n^{\prime}$ and
$\overline{n}$ results in $1^{\prime}$ and $\overline{1}$,
respectively, and for $n = 6$ the $\frac{n}{2}$-times repetition of
$n^{\prime}$ and $\overline{n}$ gives $2^{\prime}$ and
$\overline{2}$, respectively, which all forbid directional
anisotropy. In contrast, for $n = 4$ no power of $\overline{4}$ can
convert light beams with elliptical polarization to their oppositely propagating pair. For $4^{\prime}$ the same is true except for the
right- and left handed circularly polarized states which are
transformed into their counter-propagating pair.

Among the operations around an axis lying in the $xy$-plane only
$1^\prime, \overline{1}, 2$ and $\overline{2}^\prime$ can reverse
the propagation direction and preserve the polarization state of the light
beam at least for some specific (e.g. linear) polarizations. For $n = 2l$ even integers the $l$-times repetition of an $n$-fold rotation (denoted as $(n)^l$) gives $2$
and $(\overline{n}^\prime)^l$ results either in $2$ or
$\overline{2}^\prime$ depending on the parity of $l$. For $n = 4l$,
$(n^\prime)^{2l}=(\overline{n})^{2l}=2$ and for $n = 2l+1$,
$(n^\prime)^n=1^\prime$ and $(\overline{n})^n=\overline{1}$.
Therefore, the presence of $2l,
\overline{2l}^\prime,4l^\prime, \overline{4l}, (2l+1)^\prime$ or
$\overline{2l+1}$ rotations around an axis within the $xy$-plane prohibits
directional anisotropies along the $z$-axis. Rotations around arbitrary axes, which are neither parallel nor perpendicular to $z$,
cannot generally connect opposite propagation directions along the
$z$-axis.

To conclude the previous analysis, almost all symmetry operations
which reverse the direction of the light propagation retain at least
one polarization state and hereby they cannot be present in the
magnetic point groups of crystals exhibiting directional anisotropy.
The only exception is $\overline{4}$ which allows directional
anisotropy for any light polarization except for unpolarized light
where the difference cancels out.

Table \ref{table1} lists the
crystallographic magnetic point groups except those which contain
the time-reversal or spatial inversion symmetry. We indicate which symmetry groups
support directional anisotropy along certain high-symmetry axes.
In some magnetic point groups the presence of directional anisotropy can be straightforwardly demonstrated by
showing a static symmetry breaking vectorial quantity of the material which transforms under the
symmetry elements of the group in the same way as the propagation
vector of light.

\begin{table*}[!th]
\caption{Crystallographic magnetic point groups and possible configurations for directional anisotropy (DA). We omitted the point groups which contain the $1^\prime$ time reversal or $\overline{1}$ spatial inversion operation. After the
international symbol of each point group the
symmetry allowed magnetic structure (F - ferromagnetic, AF -
antiferromagnetic) is indicated. The potential directional anisotropy configurations
are listed for wave propagation parallel and perpendicular to the
principal axis as $\parallel$ to PA and $\perp$ to PA, respectively. Here i) T and t, ii)
MC, and iii) P mean the directional anisotropy effects emerging in i)
ferrotoroidic materials for propagation parallel to the toroidal
moment $\boldsymbol{T}$ and $\boldsymbol{t}$, ii) in chiral systems for light traveling along the magnetic
field, and iii) in point groups, where none of the group elements is
combined with the time reversal for propagation along the
ferroelectric polarization, respectively. X denotes cases where
directional anisotropy can also be present but it cannot be
characterized by a static vector quantity. Several examples of
multiferroic crystals belonging to the given magnetic point group
are given in the 5th and 10th columns, where $B_\alpha$ stands for an external magnetic field pointing to the $\alpha$ crystallographic direction.}

\begin{tabular}{|c|c|c|c|c||c|c|c|c|c|}
\hline
 Crystal & International & \multicolumn{2}{c|}{DA} & \multirow{2}{*}{Materials} & Crystal & International & \multicolumn{2}{c|}{DA} & \multirow{2}{*}{Materials} \\
system & notation &$\parallel$ to PA & $\perp$ to PA & & system &
notation & $\parallel$ to PA& $\perp$ to PA & \\\hline
 \multirow{2}{*}{Triclinic} & $1$ (F) & T, MC, t & T, MC, t & Ba$_2$Ni$_7$F$_{18}$\cite{Ba2Ni7F18andCsCoF4} & \multirow{8}{*}{\minitab[c]{Rhombo- \\ hedral}} & $3$ (F) & MC, P, t & X & Cu$_2$OSeO$_3$ $B_{[111]}$\cite{Cu2OSeO3}\textsuperscript{,}\footnote{FeGe\cite{FeGe}, MnSi\cite{MnSi} $B_{[111]}$; RbFe(MoO$_4$)$_2$\cite{RbFe(MoO4)2}} \\
  & $\overline{1}^\prime$ (AF) & T, t & - &  &  & $\overline{3}^\prime$ (AF) & t & X & Cr$_2$O$_3$\cite{Cr2O3} \\ \cline{1-5}
 \multirow{6}{*}{\minitab[c]{Mono- \\ clinic}} & $m$ (F) & - & T, t & BiFeO$_3$ $B_{[1\overline{1}0]}$\cite{BiFeO3}\textsuperscript{,}\footnote{Co$_3$B$_7$O$_{13}$Cl\cite{Co3B7O13Cl}} &  & $3m$ (AF) & P, t & X &  \\
  & $m^\prime$ (F) & T, t & - & Ni$_3$B$_7$O$_{13}$I\cite{Ni3B7O13I}\textsuperscript{,}\footnote{BiTeI $B_{[100]}$\cite{BiTeI}} &  & $3m^\prime$ (F) & - & X & BiTeI $B_{[001]}$\cite{BiTeI} \\
  & $2$ (F) & MC, P, t & - & 
  BaNiF$_4$\cite{BaNiF4}\textsuperscript{,}\footnote{FeGe\cite{FeGe}, MnSi\cite{MnSi} $B_{[110]}$} &  & $32$ (AF) & - & X &  \\
  & $2^\prime$ (F) & - & T, MC, t & LiCoPO$_4$\cite{LiCoPO4}\textsuperscript{,}\footnote{Cu$_2$OSeO$_3$ $B_{[110]}$\cite{Cu2OSeO3}; Co$_3$TeO$_6$\cite{Co3TeO6}; NdFe$_3$(BO$_3$)$_4$  $B_{[010]}$\cite{NdFeborat}} &  & $32^\prime$ (F) & MC, t & X &  \\
  & $2/m^\prime$ (AF) & t & - & TbOOH\cite{TbOOH}; Ba$_2$Ni$_3$F$_{10}$\cite{Ba2Ni7F18andCsCoF4} &  & $\overline{3}^\prime m$ (AF) & t & X & Gd$_2$Ti$_2$O$_7$\cite{Gd2Ti2O7} \\
  & $2^\prime/m$ (AF) & - & t & TbPO$_4$\cite{TbPO4}\textsuperscript{,}\footnote{MnPS$_3$\cite{MnPS3}; Co$_3$TeO$_6$ 17 K $<T<$ 21 K\cite{Co3TeO6}} &  & $\overline{3}^\prime m^\prime$ (AF) & - & X & Nb$_2$Mn$_4$O$_9$\cite{Nb2Mn4O9}\textsuperscript{,}\footnote{Nb$_2$Co$_4$O$_9$\cite{Nb2Mn4O9}} \\ \cline{1-5} \cline{6-10}
 \multirow{7}{*}{Rhombic} & $2mm$ (AF) & P, t & - &  & \multirow{19}{*}{\minitab[c]{Hexa- \\ gonal}} & $\overline{6}$ (F) & - & X &  \\
  & $2^\prime m^\prime m$ (F) & - & T, t & Ba$_2$CoGe$_2$O$_7$ $B_{[110]}$\cite{Kezsmarki2011}\textsuperscript{,}\footnote{CuB$_2$O$_4$ $B_{[110]}$\cite{SaitoJPSJ2008}; CdS\cite{Hopfield}, AlN, GaN, InN $B_{[100]}$\cite{wurtzite}; CaBaCo$_4$O$_7$\cite{CaBaCo4O7}; GaFeO$_3$\cite{Kubota2004};  Co$_3$B$_7$O$_{13}$Br\cite{CoB7O13Br}; KMnFeF$_6$\cite{Ba2Ni7F18andCsCoF4}} &  & $\overline{6}^\prime$ (AF) & t & - &  \\
  & $2 m^\prime m^\prime$ (F) & - & - &  &  & $6$ (F) & MC, P, t & - &  \\
  & $222$ (AF) & - & - &  &  & $6^\prime$ (AF) & - & X & ScMnO$_3$\cite{RMnO3} \\
  & $22^\prime 2^\prime$ (F) & MC, t & - & Ba$_2$CoGe$_2$O$_7$ $B_{[100]}$\cite{Bordacs2012}\textsuperscript{,}\footnote{CuB$_2$O$_4$\cite{SaitoPRL2008}, Ca$_2$CoSi$_2$O$_7$\cite{Ca2CoSi2O7}, Cu$_2$OSeO$_3$\cite{Cu2OSeO3} $B_{[100]}$; [Ru(bpy)$_2$(ppy)][MnCr(ox)] 
\cite{oxalate}}&  & $6^\prime/m$ (AF) & - & X &  \\
  & $mm m^\prime$ (AF) & - & t & LiNiPO$_4$\cite{LiNiPO4} &  & $6/m^\prime$ (AF) & t & - &  \\
  & $m^\prime m^\prime m^\prime$ (AF) & - & - &  &  & $\overline{6}m2$ (AF) & - & X &  \\ \cline{1-5}
 \multirow{18}{*}{\minitab[c]{Tetra- \\ gonal}} & $4$ (F) & MC, P, t & - & chiral metamaterial\cite{chiralmeta} &  & $\overline{6}^\prime m2^\prime$ (AF) & t & - &  \\
  & $4^\prime$ (AF) & - & - &  &  & $\overline{6}^\prime m^\prime 2$ (AF) & - & - &  \\
  & $\overline{4}$ (F) & X & - &  &  & $\overline{6}m^\prime 2^\prime$ (F) & - & X & Fe$_2$P\cite{Fe2P} \\
  & $\overline{4}^\prime$ (AF) & t & - & CsCoF$_4$\cite{Ba2Ni7F18andCsCoF4} &  & $6mm$ (AF) & P, t & - &  \\
  & $4/m^\prime$ (AF) & t & - &  &  & $6^\prime mm^\prime$ (AF) & - & X & HoMnO$_3$\cite{HoMnO3} \\
  & $4^\prime/m^\prime$ (AF) & - & - &  &  & $6m^\prime m^\prime$ (F) & - & - &  \\
  & $\overline{4}2m$ (AF) & - & - &  &  & $62(622)$ (AF) & - & - &  \\
  & $\overline{4}^\prime 2^\prime m$ (AF) & t & - &  &  & $6^\prime 2(6^\prime 22^\prime)$ (AF) & - & X &  \\
  & $\overline{4}^\prime 2m^\prime$ (AF) & - & - &  &  & $62^\prime (62^\prime 2^\prime)$ (F) & MC, t & - &  \\
  & $\overline{4}2^\prime m^\prime$ (F) & - & - &  &  & $6^\prime /mm^\prime m$ (AF) & - & X &  \\
  & $4mm$ (AF) & P, t & - &  &  & $6 /m^\prime mm$ (AF) & t & - &  \\
  & $4^\prime mm^\prime$ (AF) & - & - &  &  & $6 /m^\prime m^\prime m^\prime$ (AF) & - & - &  \\ \cline{6-10}
  & $4 m^\prime m^\prime$ (F) & - & - &  &\multirow{8}{*}{Cubic} &  & \multicolumn{2}{c|}{$\parallel$ to $3$-axis} &  \\ \cline{6-10}
  & $42(422)$ (AF) & - & - &  &  & $23$ (AF) & \multicolumn{2}{c|}{X} &  \\
  & $4^\prime 2(4^\prime 2^\prime 2)$ (AF) & - & - &  &  & $m^\prime \overline{3}$ (AF) & \multicolumn{2}{c|}{X} &  \\
  & $42^\prime (42^\prime 2^\prime )$ (F) & MC, t & - & Nd$_5$Si$_4$\cite{Nd5Si4} &  & $\overline{4}3m$ (AF) & \multicolumn{2}{c|}{X} &  \\
  & $4^\prime/m^\prime m^\prime m$ (AF) & - & - &  &  & $\overline{4}^\prime 3m^\prime$ (AF) & \multicolumn{2}{c|}{-} &  \\
  & $4/m^\prime mm$ (AF) & t & - &  &  & $43(432)$ (AF) & \multicolumn{2}{c|}{-} &  \\
  & $4/m^\prime m^\prime m^\prime$ (AF) & - & - &  &  & $4^\prime 3(4^\prime 32)$ (AF) & \multicolumn{2}{c|}{X} &  \\
  &  &  &  &  &  & $m^\prime \overline{3}m$ (AF) & \multicolumn{2}{c|}{X} &  \\
  &  &  &  &  &  & $m^\prime \overline{3}m^\prime$ (AF) & \multicolumn{2}{c|}{-} &  \\     \hline 
\end{tabular}
\label{table1}
\end{table*}

A static toroidal moment ($\boldsymbol{T}$ or $\boldsymbol{t}$) transforms under all possible
symmetry operations identically to the wave vector since the
latter is the cross product of the electric and
magnetic components of light. Therefore, systems with finite
toroidal moment vector allow directional anisotropy for beams propagating
parallel or antiparallel to this vector and the corresponding effect is termed as the OME.
The magnetic point groups compatible with $\boldsymbol{T}$ -- denoted by T in Table \ref{table1} -- are subgroups of the
symmetry group of a perpendicular magnetic and electric field
vector, $2^\prime m m^\prime$, where the electric field is parallel to $2^\prime$, magnetic field is perpendicular to the mirror plane $m$ and $\boldsymbol{T}$ points perpendicular to the plane $m^\prime$. In the notation of the magnetic point
groups we follow the international convention, where $m$ stands for
mirror plane symmetry. The toroidal moment $\boldsymbol{t}$ can arise without the presence of electric polarization and magnetization, thus, $\boldsymbol{t}$ is compatible with higher symmetry point groups than $\boldsymbol{T}$. Indeed, $\boldsymbol{t}$ can exist the same point groups as a general time-reversal odd polar vector. In the case of crystals these are the subgroups of $n/m^\prime mm$ and  $\overline{n}^\prime 2^\prime m$ where $\boldsymbol{t}$ is parallel to the principal axis and $n = 4$ or $6$. Since a spontaneous, non-dissipative electric current\cite{Ascher} has the same symmetry properties as $\boldsymbol{t}$, they can be present in the same 31 crystallographic magnetic point groups. These groups are labeled with t in Table \ref{table1}.

Chiral systems only possess proper rotational symmetries, which may
be combined with the reversal of time. All of these operations transform
the axial vector of the magnetic field (or spontaneous magnetization) and the polar wave vector in the same way. Therefore, in chiral
systems with a time-reversal odd axial vector order parameter such as magnetization light propagation
along or opposite to this vector cannot be connected by symmetry. The corresponding directional anisotropy is the so-called MChD and is indicated by MC in Table \ref{table1}.

Additionally to the well-known OME and MChD, we found if a polar magnetic point group contains only symmetry elements
which are not combined with time-reversal then these operations
transform the vector of electric polarization or temperature gradient, both even in time reversal, in the
same way as the wave vector being odd in time reversal. In this case the
presence of a static polarization/temperature gradient makes the wave propagation along
or opposite to it inequivalent. This potential directional
anisotropy effect, denoted by P in Table \ref{table1}, has not been studied either theoretically or
experimentally so far.

Please note that whenever symmetry permits directional anisotropy of either MC- or P-type, t-type directional effect can also emerge. Nevertheless, the distinction between these cases is meaningful as the physical origin of these directional effects can be different. Besides these cases we found additional situations when the symmetry condition for directional anisotropy cannot be related to any static vectorial quantity which transforms in the same way as the wave vector. These point groups are marked with an 
X in Table \ref{table1}.


Except for the magnetic point group $\overline{4}$, all the
cases of directional anisotropy listed in Table
\ref{table1} also have validity for other processes than light
propagation. Symmetry elements of these groups cannot reverse wave propagation and, thus, allow directional anisotropy for propagation of electrons, neutrons, acoustic- and
spin-waves, and in electric or heat conduction. In some other
groups directional anisotropy is conceivable but only for polarized light, spin polarized neutron and electron beams or any transversal wave. These cases can be determined
by systematical checking whether the propagation-reversing
operations can preserve the polarization state. Specific to light propagation, we found that $\overline{4}$ is the only additional point group supporting directional optical anisotropy.

Another approach to study directional optical anisotropy is the symmetry analysis of the optical response functions. Due to its simultaneous time-reversal and spatial inversion breaking nature, optical directional anisotropy can only be produced by the bilinear magneto-electric tensor
when dipole approximation appropriately describes the light-matter interaction. The non-vanishing elements of this tensor can be identified by the symmetry analysis of the given material. For the static magneto-electric tensor the result of such analysis based on magnetic point groups can be found in the literature\cite{HandbookMagMatVol19}. While the static magneto-electric tensor can only contain time-reversal odd elements, the dynamic one may have time reversal even part as well. Therefore, the symmetry dictated form of the dynamic magneto-electric tensor needs to be included in the constitutive relations in order to solve the Maxwell's equations and check the presence of directional anisotropy. This approach is specific to optics and its validity is restricted to the dipole approximation of light-matter interaction. Considering compounds with cubic magnetic point groups, in dipole order no directional anisotropies can be present since the magneto-electric tensor is proportional to the unit tensor. However, taking into account higher order terms described e.g. by electric quadrupole tensor, the symmetry allowed directional anisotropy along the threefold axis shown in Table \ref{table1} can be reproduced. Nevertheless, we think that our approach provides a more general framework to address the question of directional anisotropies in crystalline solids.

In conclusion, possible direction-dependent optical configurations were
determined using symmetry considerations. Besides the conventional
OME and MChD a third potential directional anisotropy effect was
predicted when the wave propagation is parallel to the electric
polarization in materials whose symmetry groups do not contain any symmetry operation involving time reversal. With one exception, all
the listed configurations allow non-reciprocity for any transport
phenomenon, particle or wave propagation. We identified each
crystallographic magnetic point group which can host directional
anisotropies providing a key for the systematic investigation of
candidate materials. We also propose a large collection of materials where these effects are expected to emerge.

We thank G. Kriza, Y. Tokura, L. De\'ak, T. F\"ul\"op and B. D\'ora for valuable discussions.
This project was supported by Hungarian Research Funds OTKA PD75615, T\'AMOP-4.2.2.B-10/1-2010-0009.

\end{document}